\documentclass[journal]{IEEEtran}
\ifCLASSINFOpdf
  \usepackage[pdftex]{graphicx}
 \graphicspath{{C:\FreqDiversity_paper\arxiv_FD}}
  \DeclareGraphicsExtensions{.pdf,.jpeg,.png}
\else
\fi
\usepackage{amsmath, bm}
\usepackage{fixmath}
\usepackage{textcomp}
\usepackage{amssymb}
\usepackage{gensymb}

\usepackage{subfig}

\begin{document}

%
\title{Inter Pulse Frequency Diversity System for Second Trip Suppression and Retrieval in a Weather Radar}
%
%
%

\author{V~Chandrasekar,~\IEEEmembership{Fellow,~IEEE},
		Mohit~Kumar,~\IEEEmembership{Student Member,~IEEE}}

%
%

\markboth{}%
{Shell \MakeLowercase{\textit{et al.}}: Bare Demo of IEEEtran.cls for IEEE Journals}
%



\maketitle

\begin{abstract}
This paper develops the use of Inter-pulse frequency diversity scheme for a weather Radar system. It establishes the performance of frequency diversity technique comparing it with other inter-pulse schemes for weather radar systems. Inter-pulse coding is widely used for second trip suppression or cross-polarization isolation. Here, a new inter-pulse scheme is discussed taking advantage of frequency diverse waveforms. The simulations and test of performance, is evaluated keeping in mind NASA dual-frequency, dual-polarization, Doppler radar (D3R). A new method is described to recover velocity and spectral width due to incoherence in samples from change of frequency pulse to pulse. This technique can recover the weather radar moments over a much higher dynamic range of the other trip contamination as compared with the popular systematic phase code, for second trip suppression and retrieval. 
\end{abstract}


%
\IEEEpeerreviewmaketitle

\section{Introduction}
%
%
%
%
\IEEEPARstart{T}{he} phase coding method tags the transmit waveform with a phase code and the same could be demodulated to retrieve the pulse parameters (amidst interference from different pulses like second trips or cross polar coupling) in a weather radar system. The phase code can be a single tag used for a pulse, (as in case of inter-pulse coding) to separate out multiple trips. This can essentially aid in suppressing the effects of other trips on the one intended, and also for the recovery of these multiple trips under certain restrictions. The suppression effect of the phase code could be observed over the full cycle of integration, which is the coherent processing interval to generate polarimetric moments from raw data.\par
In a pulse doppler weather radar, the pulses are transmitted at the pulse repetition interval of $\tau$ sec, the maximum unambiguous range is given by $r_{unb} = c\tau/2$ and the maximum unambiguous velocity is given by $v_{unb} = \lambda/4\tau$, where $\lambda$ is the wavelength of operation. Hence both $r_{unb}$ and  $v_{unb}$ are inversely proportional to each other.
Due to this, unambiguous range and velocity cannot be simultaneously optimized. If one is increased, the other one is inversely affected. This is termed as range-doppler dilemma \cite{Bharadwaj2007}. A high Pulse repetition frequency (PRF) radar system would be able to resolve the doppler frequency better in a wider range but the unambiguous range would suffer, whereas a low PRF system, would detect weather upto a higher range but the velocity would be folded in the doppler domain. For a medium PRF system, it would be difficult to resolve both range and velocity. This problem is more prominent for weather radar systems as the scatterers are continuous and distributed throughout the beam illumination volume, with wide dynamic range \cite{Bringi2001}.\par

Different phase coding schemes exist in literature to overcome this basic limitation and retrieve polarimetric moments from the second, third and so on, trip echoes (after the unambiguous range) for dual polarimetric weather radar. These codes can be broadly classified into intra-pulse (phase changes on a sub-pulse basis, \cite{kumar2019intrapulse}) and inter-pulse (phase change on a pulse basis) code. The random phase code and systematic phase code, are examples of Inter-pulse phase code, popular to retrieve first and overlaid second trip \cite{Zrnic1999}.

Staggered or dual PRF techniques are some another methods \cite{Torres2017}, \cite{Cho2005}, generally used to improve upon $r_{unb}$ and $v_{unb}$. Particularly, they are very effective in increasing the range of $v_{unb}$. Typically, it is accomplished through the use of PRF diversity by playing two prf's one after the other or in batches, but the fundamental concept remain the same. That is, by observing the same volume with different prf's. However, it consumes scan time and usually uses $N$ times the time required for the uniform prf to play, $N$ being the number of stagger pulses. Hence it is expensive on the radar time.\par

Inter-pulse codes have been explored extensively in weather radar community for second trip suppression \cite{Zrnic1999} and for orthogonal channel coding in \cite{Chandrasekar2009} for dual-polarization weather radars. The retrieval of moments for the first and subsequent trips, is based upon spectral processing of weather echoes in batches of coherent interval (which in turn depends upon antenna rotation rate and weather de-correlation time). The orthogonal property of the code, for different trips, is achieved over the coherent interval as the second trips get modulated by cyclic shifts of the phase code and the other trips by multiple cyclic shifts. The orthogonal nature is achieved by having these cyclic shifts uncorrelated to each other. The systematic codes use derivatives of Chu codes which have deep nulls in the correlation function for time delayed versions of itself. However, the rejection of echoes from the trips of non-interest, are a function of spectral width and would degrade in case of multi-modal distribution, wider spectral width and phase noise.\par

\begin{figure}[!t]
	\centering
	\includegraphics[width=3in]{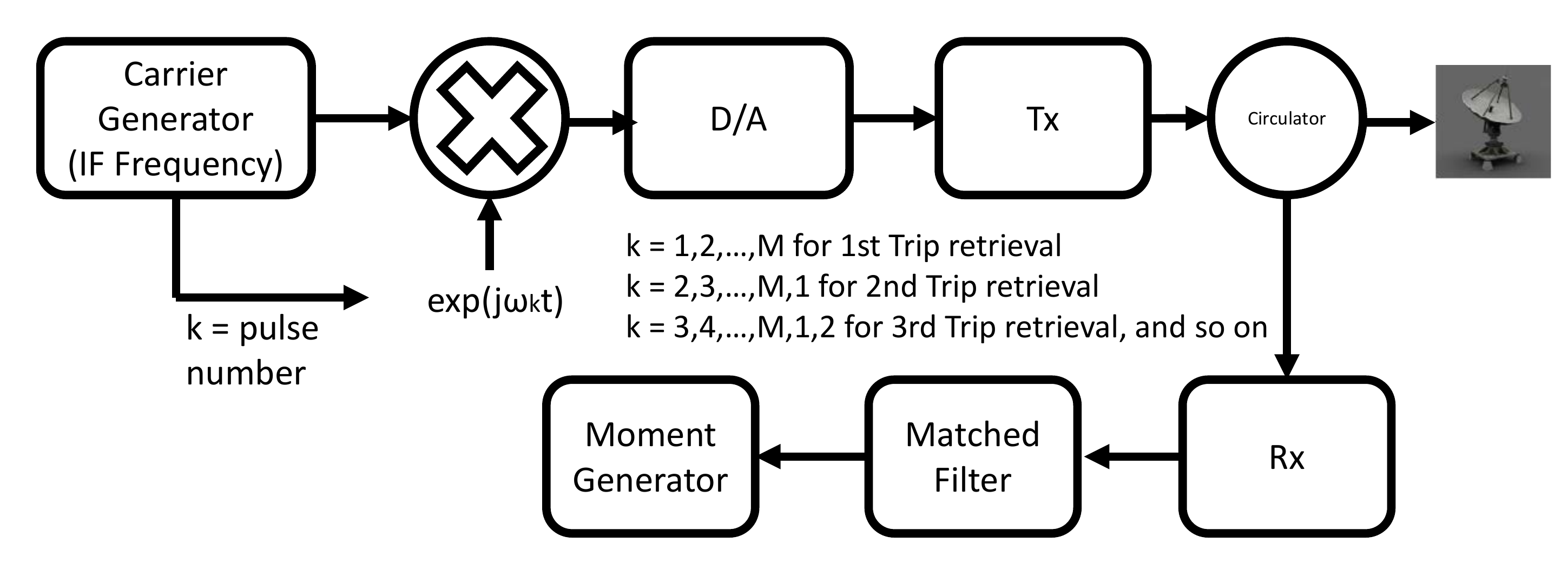}
	\caption{The system architecture which makes use of frequency diversity scheme for multi-trip retrieval. The pulse to pulse frequency change is done at the down-converter.}
	\label{fig_blockDia}
\end{figure}

In this paper, we introduce a novel frequency diverse inter-pulse system and its implementation on NASA D3R weather radar. In this scheme, we change the IF frequency from pulse to pulse, for example, for second trip suppression, we use two frequencies, $\omega_{1}$ and $\omega_{2}$ for alternate pulses and re-synchronize with these on digital down-converter to remove second trip. As will be shown, it gives excellent performance, if the frequencies are far apart. We also discuss a novel method to retrieve velocity and spectrum width, for the first and second trips. Since the two frequencies are far apart, the data from adjacent pulses (modulated with $\omega_{1}$ and $\omega_{2}$) are uncorrelated, and a method is highlighted which can recover moments, from a batch of coherent processing time (128 pulses, in case of D3R), under such circumstances. The improvement in the second trip recovery region, based on range of $\{P_{1}/P_{2}\}$, where $P_{1}$ is first trip power and $P_{2}$ is second trip power, is immense. A block diagram of this scheme is shown in figure \ref{fig_blockDia}. The carrier generator output two different frequencies, $\omega_{1}$ and $\omega_{2}$, on odd and even pulse, respectively. If the sequence at down-converter is $\omega_{1}$ and $\omega_{2}$ for odd and even pulse, we would be able to recover first trip. However, if the sequence is $\omega_{2}$ and $\omega_{1}$, then we would be recovering second trip parameters. \par

In contrast to the scheme of dual-pulse, dual-frequency technique for range and velocity ambiguity mitigation highlighted in \cite{Torres2010}, our technique is working towards the suppression of second trip echoes. The authors in \cite{Torres2010} have tried to improve on range and velocity measurements to a much higher range and higher velocities by making use of uncorrelated frequencies in two different channels. With this, they have arrived at an algorithm to derive velocity which becomes independent of base PRT. However, this paper doesn't address the problem of second trip, which might still be present in case the weather event is still happening beyond the increased unambiguous range (due to the processing of dual channels). \par

Many orthogonal polyphase coded systems have been proposed in literature, as in \cite{Haohe2009}, \cite{Song2016}, \cite{Deng2004} and \cite{Griep1995}. But it is very difficult to get an isolation between different polyphase codes $> 40dB$. The frequency diversity scheme, proposed in this paper, is meant to give a higher level of isolation that is possible with the polyphase coded systems. The limit on peak auto-correlation and cross-correlation sidelobe, of sequences have been brought out in \cite{Sarwate1979} and \cite{Welch1974}, therefore there is a need to explore other techniques. Also, this frequency diversity scheme, in general, can be applied to obtain simultaneous co-polar and cross-polar echoes, in case of dual polarization weather radar leading to instantaneous radar polarimetry \cite{Howard2007}, \cite{Wang2010}. However, this is not the focus of this paper. Additionally, orthogonality requirements in case of MIMO systems \cite{Stoica2008} can be achieved via this scheme. Moreover, the same logic holds good for CDMA based communication as well \cite{Liu1995}. The latest efforts to gain on orthogonality between waveforms have been towards the concept of zero-autocorrelation and cross-correlation sidelobes (perfect sequences) offered by Golay type waveforms. However, they display perfect orthogonal behavior over very limited Doppler. A lot of research is currently happening in making these sequences more doppler tolerant \cite{Tang2014}, \cite{Pezeshki2008}, \cite{Nguyen2016} and \cite{Yang 2007}.\par
This paper is organized as follows: A detailed description of the frequency diversity scheme and the Chu phase coded inter-pulse scheme are in section \ref{section1}. Section \ref{section2} analyze the effect of frequency change at IF level, on other dual-polarization moments. Section \ref{section3} presents the D3R weather radar data on which this scheme was implemented and tested, followed by Conclusion in section \ref{section5}.

\section{Inter-Pulse Chirp Waveforms} \label{section1}

\subsection{Generalized Chu codes for second trip mitigation and retrieval:}
The most popular inter-pulse code, are the systematic phase codes, commonly known as SZ Code, for the retrieval of parameters of overlaid echoes. The SZ code, which are the derivatives of the Chu code, with good correlation properties, have better estimation accuracies over a wide dynamic range of the overlay power and spectral width. In this paper, the performance of Chu code is analyzed for second trip suppression and retrieval. Then we compare it with the frequency diversity scheme. Under wide spectral width, the systematic phase code, which rely on replicating the other trip's spectrum multiple times, looks more whitened. Another inter-pulse code which introduces random phase on pulses (known as random phase code), attempts to whiten the weaker trip, while it coheres the stronger trip signal, whereas the SZ code produces replicas of the weaker signal spectrum. If a notch filter is used in random phase code, it also notches out some part of the other trip spectrum, which cannot be recovered later on, thus producing a self noise \cite{Frush2002}.
However, in case of SZ code, even after the stronger trip is notched out completely, by using a wider filter width and retaining two replicas of the weaker trip, we can still recover the velocity and spectrum width of the weaker trip.  The phase difference between two replicas is sufficient to estimate velocity. While notching out, we need to remember that the Gaussian spectrum of the weather gets broadened out due ground clutter and phase noise etc. Thus, to notch out the stronger trip, we generally need a much wider notch filter than what Gaussian spectrum width would require to be. \par
Another reason for the generation of self noise is due to overlap of the spectral replicas. If the spectrum of the signal is well contained within $M/8$ spectral coefficients (for SZ(8/64)), then there would be less overlap region between successive replicas and better estimation can be performed. But in case of wider spectral width, the overlap region can pose a constraint for the recovery region of velocity and spectral width. In such a case, probably SZ(4/64) code with $n=4$, will provide a better separation but then, it will allow for much lesser notch width, so as to retain a minimum of two spectral replicas. Additionally, the phase noise leads to broadening of the spectrum, which can be linked to phase noise of the coherent oscillator used to synchronize various sub-systems of the radar.\par
The SZ code, simulated here, is designed in accordance with equation \eqref{eq_1} with $n=8$ and $M=64$. 

\begin{equation}\label{eq_1}
\begin{aligned}
c_{k} &= exp(-j \varSigma_{n=1}^{M}(n \pi m^{2}/M))\\
& k = 0, 1, 2, ..., M
\end{aligned}		
\end{equation}

\begin{figure}[!t]
	\centering
	\includegraphics[width=3in]{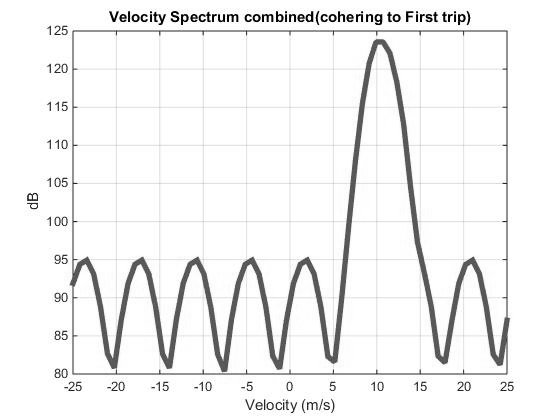}
	\caption{The modulation code spreads out the power of second trip to 8 replicas in the spectral domain.}
	\label{fig_chu1}
\end{figure}

The cyclic version of this code (known as modulation code) generate eight replicas of the cohered trip in the Nyquist interval along with the original spectrum of the other trip. We carried out simulation of weather echoes, where the first trip has the parameters: $v_{1} = 10m/s$, $w_{1} = 1m/s$ and $\rho_{hv} = 0.995$ and the second trip has the same parameters except velocity is $v_{1} = -5m/s$.  The simulation was carried out with the method highlighted in \cite{GALATI199517}. This was done to see the effectiveness of systematic phase code and later use it to compare with frequency diversity scheme. The simulation was carried out at S-band with $PRF = 1.2KHz$. On the receive, when the first trip is being retrieved, the spectrum consists of first trip specturm and the second trip spectrum getting replicated 8 times (in case of SZ(8/64)), with the power of second trip getting spread out in all these replicas, as shown in figure \ref{fig_chu1}.

\begin{figure}[!t]
	\centering
	\includegraphics[width=3in]{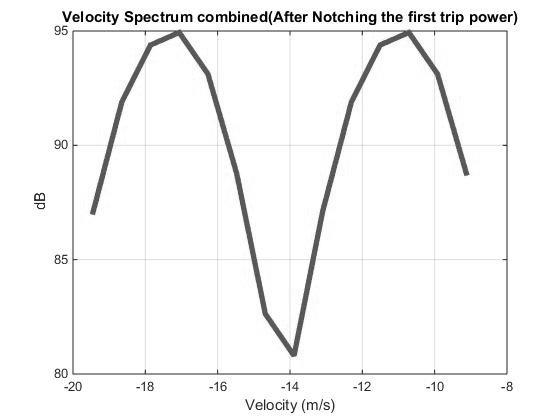}
	\caption{The retained spectrum after the application of notch filter with normalized width of 0.75.}
	\label{fig_chu2}
\end{figure}

After estimation of first trip parameters, the spectrum is notch filtered, so as to remove the power of first trip, and second trip parameters can be estimated. If a rectangular window is used for truncating the signal (after 128 pulses), then the stronger trip will be contaminating the weaker trip spectrum through its spectral sidelobes, and the dynamic range,  $\{P_{1}/P_{2}\}$, where the weaker trip could be retrieved, will reduce. The reduction in spectral sidelobe leakage is important to gain on this dynamic range. However, use of other window functions will leads to loss of SNR, which in turn, means a reduction in number of independent samples. We used Hann window for simulation, which has SNR loss of 4.2 dB, but the spectral dynamic range gets increased to 90dB. After the notching process, at least two replicas need to be retained for velocity and spectral width estimate. The spectrum after notch process, is shown in figure \ref{fig_chu2}.

\begin{figure}[!t]
	\centering
	\includegraphics[width=3in]{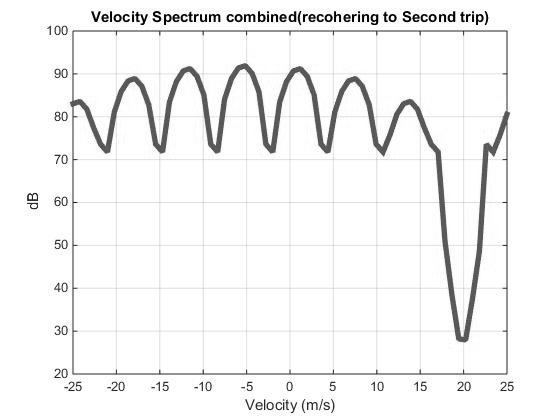}
	\caption{The remaining spectrum is re-cohered for the second trip with its six sidebands present in the spectrum.}
	\label{fig_chu3}
\end{figure}

The remaining signal is re-cohered for the second trip, and the resulting signal has six symmetrical sidebands centered at the mean velocity of the second trip. This is shown in figure \ref{fig_chu3}.

\subsection{Effect of Phase Noise on Chu Codes:}
The phase noise of the system is a major limiting factor for the dynamic range of $P_{1}/P_{2}$, where parameters of both the strong and the weak trip, can be recovered, using a inter-pulse Chu code. Phase noise leads to broadening of spectrum. The overall phase noise is dominated by the phase noise of the basic oscillator, from which all the other clocks are derived. Single-side band phase noise is usually measured in a 1 Hz bandwidth, and is defined as the carrier power at an offset with respect to the power measured at 1 Hz Bandwidth. More precisely, phase noise can be defined as the ratio of noise in a 1 Hz Bandwidth to the signal power at the center frequency.\par
The equivalent noise jitter can be obtained by integrating the phase noise curve over the receiver bandwidth. It is equivalent to:

\begin{equation}
\text{RMS Phase Jitter (in radians)} = \sqrt{2 \times 10^{A/10}}
\end{equation}
where $A$ is the area under the phase noise curve. 

\begin{figure}[!t]
	\centering
	\includegraphics[width=3in]{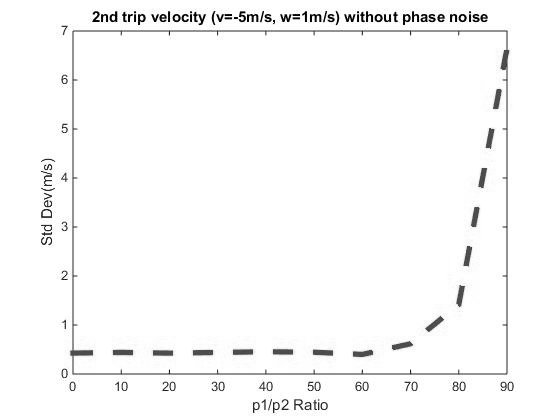}
	\caption{The dynamic range of $P_{1}/P_{2}$, in which the weaker trip velocity can be recovered, without phase noise.}
	\label{fig_WOPN}
\end{figure}

\begin{figure}[!t]
	\centering
	\includegraphics[width=3in]{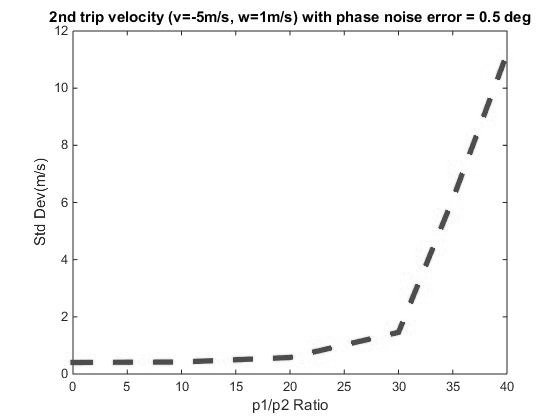}
	\caption{The dynamic range of $P_{1}/P_{2}$, in which the weaker trip velocity can be recovered, with phase noise.}
	\label{fig_WPN}
\end{figure}

 It has been shown in \cite{Zrnic1999} that, if there is no phase noise and Hann window function is used, the limit on retrieval of velocity is about 90dB for $P_{1}/P_{2}$ ratio, for spectral width less than 4 m/s. But it drastically reduces to 60dB, if rms jitter is of the order of 0.2 deg rms, and further reduces to 40dB in case of 0.5 deg rms phase jitter. This has been confirmed through our simulation also, that the accuracy of the velocity retrieval drastically reduces, under phase noise conditions. The dynamic range of $P_{1}/P_{2}$, in which the weaker trip velocity can be recovered, with and without phase noise, is shown in figures \ref{fig_WOPN} and \ref{fig_WPN} respectively.

\subsection{Frequency Diverse Chirp Waveforms:}
\begin{figure}[!t]
	\centering
	\includegraphics[width=3in]{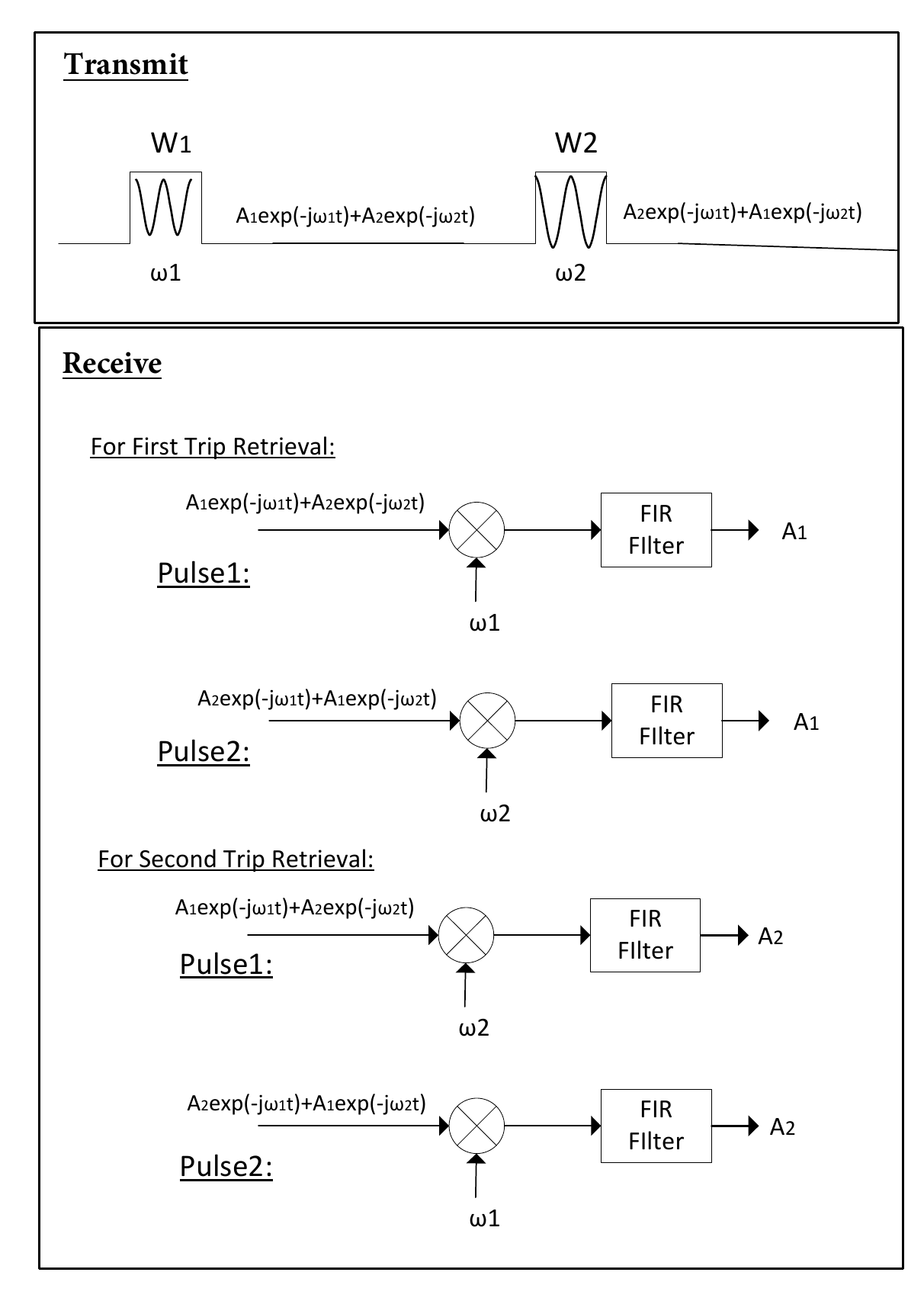}
	\caption{The use of frequency diversity at IF frequency with NCO being switched from $\omega_{1}$ to $\omega_{2}$ and back, from pulse to pulse.}
	\label{fig_freqDiv}
\end{figure}

Modern day systems with higher computation power and embedded with FPGAs (Field Programmable Gate Arrays) are capable of complex signal processing architecture and speed like never before. The digital Receiver system in D3R, is capable of switching IF (Intermediate Frequency) on a pulse by pulse basis (\cite{Kumar8517944}). This feature has been utilized here in obtaining frequency diversity at IF frequency. The main factor that would limit the amount of second trip suppression possible, would be the IF filter, which is implemented digitally in FPGA (based on its stop band suppression and roll off in frequency domain). That would decide, how we should be doing the frequency planning for obtaining adequate removal of the undesired trip echo. This has been explained with the help of figure \ref{fig_freqDiv}.

Typically, when the transition of the filter from passband to stopband is steep, it would require high number of digital MAC (multiply-accumulate) units. But with the advent of high processing power and FPGA nodes, optimized for DSP application, we can easily obtain very sharp roll-off filter working in real time. The analog filter before A/D converter must be wide enough to accommodate both $\omega_{1}$ and $\omega_{2}$, in addition to the transition band of the digital filter. That is why, a sharper roll off is important, so that we save on bandwidth. Finally, it is difficult to obtain high analog bandwidth stages, because of spurious and inter-modulation products in the mixing process, which may appear in-band. This also leads to reduction in spurious-free dynamic range (SFDR). \par
In figure \ref{fig_freqDiv}, the inter-pulse IF change of the transmit waveform is shown, where $\textbf{W}_{1}$ and $\textbf{W}_{2}$ can, in general, be an orthogonal pair. However, in our design, $\textbf{W}_{1} = \textbf{W}_{2}$, and is a LFM waveform (Linear Frequency Modulated) with pulse width = $20\mu$s and bandwidth = 1 MHz and the pulse repetition frequency is 0.5 KHz. The $\omega_{1}$ and $\omega_{2}$ will be selected based on the digital filter characteristics and will be dealt in more detail, later in this section. If $A_{1}$ corresponds to first trip and $A_{2}$ denotes the amplitude of second trip echo, then on the receive IF processing when pulse 1 goes through the final IF mixing stage, the output at mixer out port will be:

\begin{multline}
O_{1} = A_{1} + A_{1}exp(-j(2\omega_{1})t) + A_{2}exp(-j(\omega_{2}-\omega_{1})t) \\
+ A_{2}exp(-j(\omega_{2}+\omega_{1})t)
\end{multline} 

For first trip processing, the nearest component to be filtered out is $A_{2}exp(-j(\omega_{2}-\omega_{1})t)$ and the first trip information is in $A_{1}$, has to retained. Similarly, for the other pulse as well, we will retain $A_{1}$ and filter out the nearest second trip component, $A_{2}exp(-j(\omega_{1}-\omega_{2})t)$. Later, we coherently integrate for many such pulse sets and retrieve first trip information. The process for second trip retrieval is very similar to this.\par

\begin{figure}[!t]
	\centering
	\includegraphics[width=3.6in]{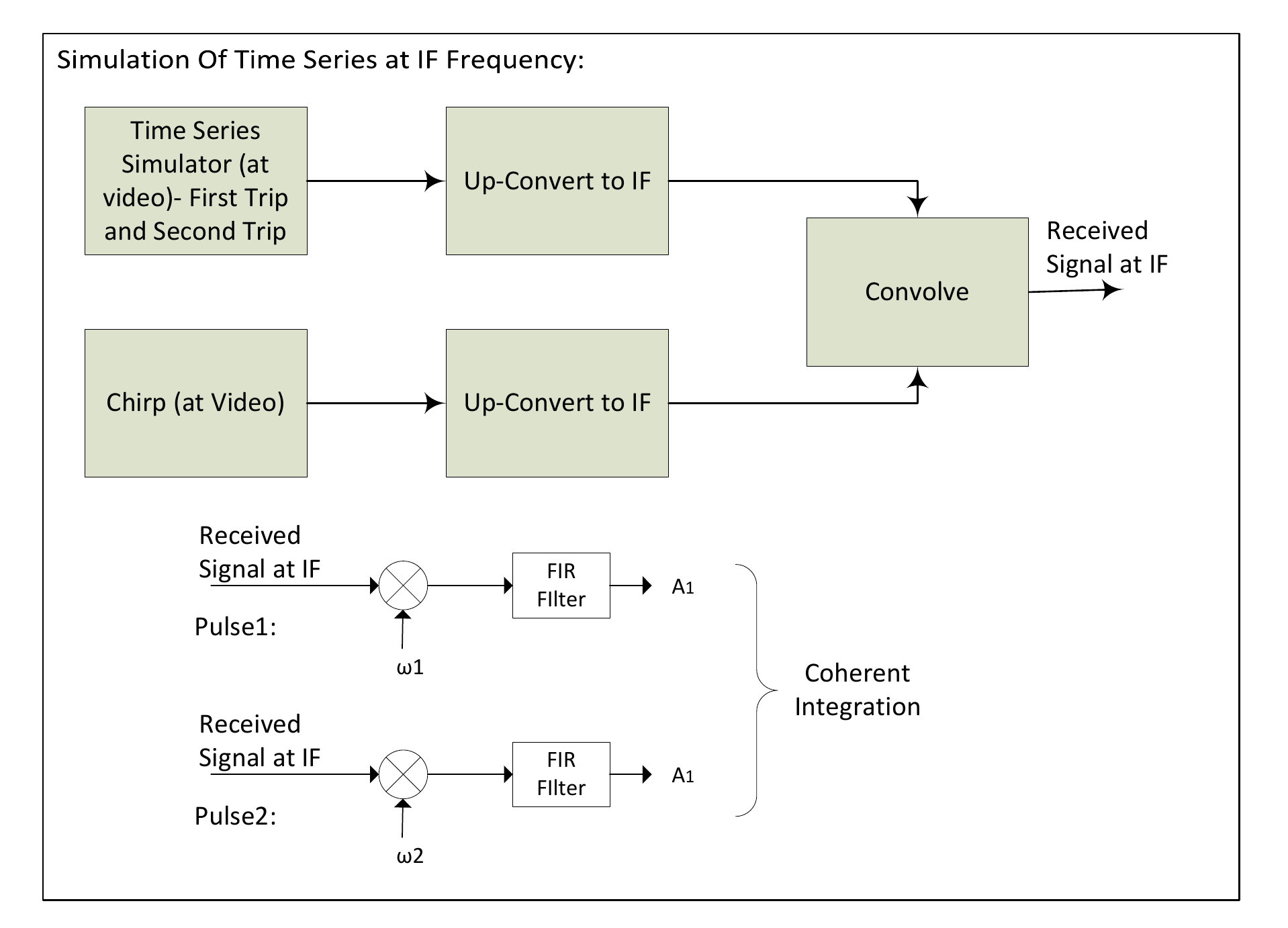}
	\caption{The time series simulation at IF Frequency.}
	\label{fig_IFTime}
\end{figure}

To demonstrate the frequency diversity scheme, we start with simulation of weather echoes with parameters:  $v_{1} = 10m/s$,$w_{1} = 1m/s$ and $\rho_{hv} = 0.995$ and the second trip has the same spectral width and co-polar correlation coefficient, except velocity is $v_{2} = -5m/s$. This simulated time series will be modulated with  $\omega_{1}$ for the first pulse and  $\omega_{2}$ for the next pulse in a frame.   The frame is a basic unit of two pulses, which repeats. The pulse wise returns are then filtered by a digital filter at baseband and then after pulse compression, the power of echo signal is calculated. Based on this, we would get the dynamic range of values, $P_{1}/P_{2}$, where the parameter retrievals are within acceptable range (based on measured bias and standard deviation). This process of time series simulation is shown in figure \ref{fig_IFTime}.\par

\begin{figure}[!t]
	\centering
	\includegraphics[width=3in]{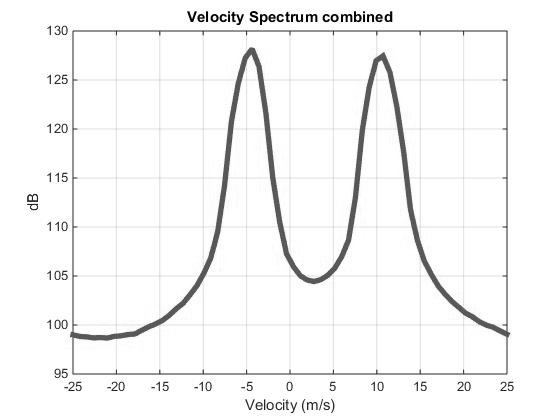}
	\caption{Both the first and the second trip echoes are generated with equal power such that  $p1/p2 = 0$dB and with parameters: $v_{1} = 10m/s$,$w_{1} = 1m/s$ and $\rho_{hv} = 0.995$ while the second trip has the same parameters, except velocity of $v_{2} = -5m/s$. }
	\label{fig_IFfig1}
\end{figure}

\begin{figure}[!t]
	\centering
	\includegraphics[width=3.6in]{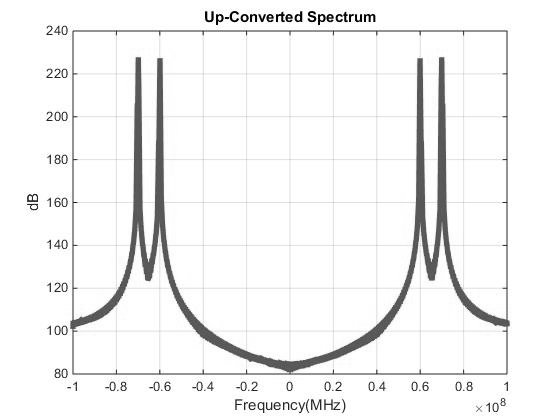}
	\caption{The Spectrum of Up-Converted First and Second trip echoes with  $\omega_{1} = $60MHz and $\omega_{2} = $70MHz. }
	\label{fig_IFfig2}
\end{figure}

\begin{figure*}[!t]
	\centering
	\includegraphics[width=7in]{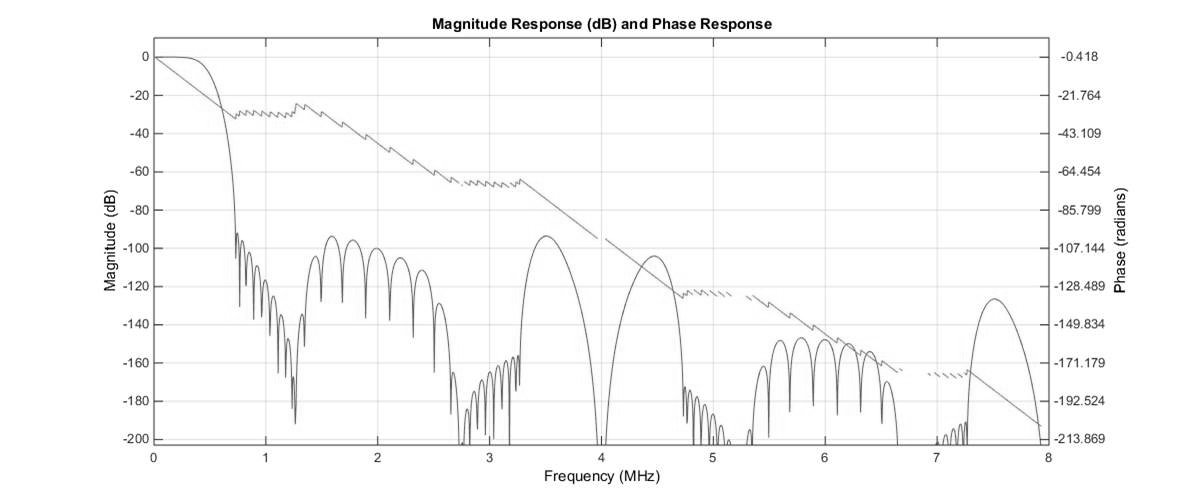}
	\caption{The filter response of the down-converter stage. }
	\label{fig_IFfig3}
\end{figure*}

Figure \ref{fig_IFfig1} shows both the first and second trips, generated with equal power, such that $P_{1}/P_{2} = 0$dB, and both of them have phase jitter of 0.5 deg rms. The waveforms are up-converted to IF frequency. The odd pulse first trip, is up-converted to $\omega_{1}$ and combined with second trip echo (up-converted to $\omega_{2}$). In the same way, the even pulse first trip, is up-converted to $\omega_{2}$ and combined with second trip echo (being up-converted to $\omega_{1}$). The spectrum after this process, where, $\omega_{1}$ = 60MHz and $\omega_{2}$ = 70MHz, is shown in figure \ref{fig_IFfig2}. Next, the time series is down-converted with these IF frequencies. In this whole process, if we have one down-converter and pulse compressor channel, we can either retrieve first trip by switching the Numerically Controlled Oscillator (NCO) to a sequence 1: $\omega_{1}$, $\omega_{2}$ for a frame of two pulses, based on odd or even pulse. For retrieval of second trip, we would need to switch to a sequence 2: $\omega_{2}$, $\omega_{1}$ for the two pulse frame. But, if we have two channels of down-converter and pulse compressor system, then we can do this in parallel, by programming NCO to sequence 1 or 2, depending upon the trip to be retrieved (in respective channels). This would consume double the resources in an FPGA based system, compared with a single channel configuration. Another approach could be, to use alternate pulse-pair frames for first trip, and in-between frame for second trip. In such a case, the sequence of NCO would be : $\omega_{1}$, $\omega_{2}$, $\omega_{2}$ and $\omega_{1}$ for a frame of four pulses. This scheme would save on the resource and computational complexity. The down-converter has been set up here, to use frequency conversion followed by filtering and decimation stage. These stages are composed of FIR and CIC filters. The overall frequency response of the down-converter stage is set to passband ripple of 0.2dB and stop-band attenuation of 80dB. The amplitude and phase response of the filtering stage is shown in figure \ref{fig_IFfig3}.

\begin{figure}[!t]
	\centering
	\includegraphics[width=3.6in]{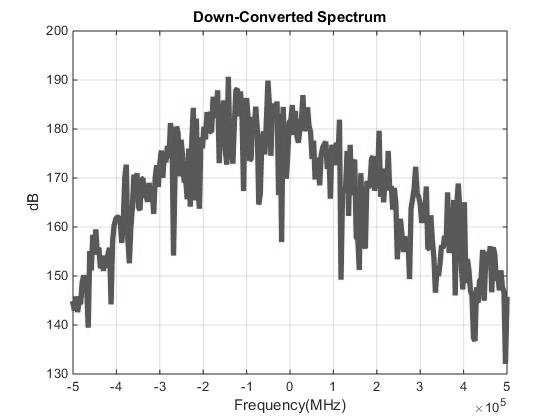}
	\caption{The Spectrum after down-converter stage (cohering to second trip). }
	\label{fig_IFfig4}
\end{figure}

\begin{figure}[!t]
	\centering
	\includegraphics[width=3.6in]{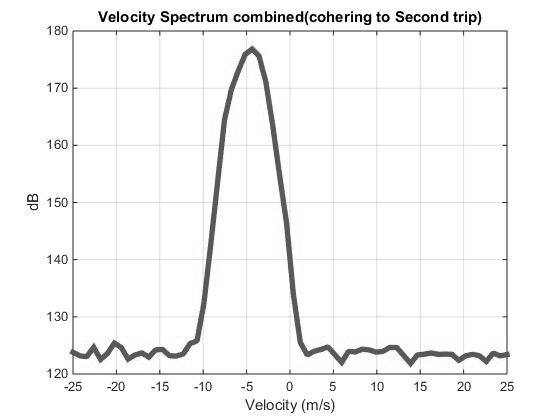}
	\caption{The Velocity Spectrum of the second trip, recovered after frequency planning of $\omega_{1}$ and $\omega_{2}$.}
	\label{fig_IFfig5}
\end{figure}

The spectrum after down-conversion and filtering stage, for the sequence (for NCO) set for the second trip retrieval is shown in figure \ref{fig_IFfig4}. The bandwidth of chirp is 0.5 MHz. After down-converter, the data goes through pulse compression stage. As a final computation, we reconstruct the velocity plot of second trip, with a power ratio $P_{1}/P_{2} = 0$dB, over the number of integration pulses (N = 64 here). This is shown in figure \ref{fig_IFfig5}. More detailed analysis of the velocity and spectral width reconstruction process, is given in next section.

The noise floor is dominated by phase noise of the echo signal received, which negatively impacts the dynamic range of $P_{1}/P_{2}$, that can be reconstructed. This was also highlighted before, during the recovery of second trip echoes, using Chu inter-pulse codes. It was shown that the second trip could be recovered for the power ratio ($P_{1}/P_{2}$) upto 40dB, under 0.5 deg rms phase jitter. Under similar condition of phase noise (jitter), the frequency diversity scheme, developed in this paper, can recover second trip for the power ratio spanning 60dB, an improvement of 20dB over the Chu inter-pulse code. This is one of the major achievement of this work. This has been substantiated here, with simulation, and the mean bias and standard deviation of second trip velocity as a function of power ratios, for frequency diverse scheme, is shown in the figures \ref{fig_IFfig6} and \ref{fig_IFfig7} below.

\begin{figure}[!t]
	\centering
	\includegraphics[width=3.6in]{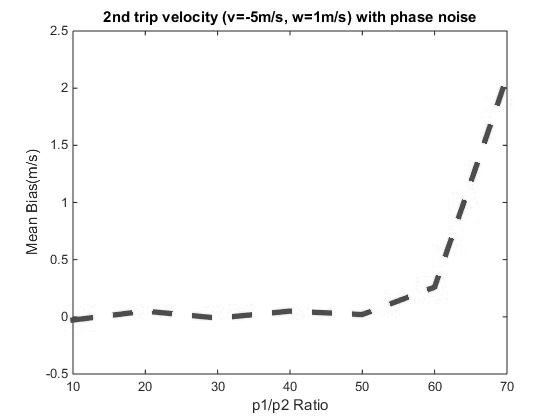}
	\caption{The Mean Bias in the second trip velocity, after frequency planning of $\omega_{1}$ and $\omega_{2}$. }
	\label{fig_IFfig6}
\end{figure}

\begin{figure}[!t]
	\centering
	\includegraphics[width=3.6in]{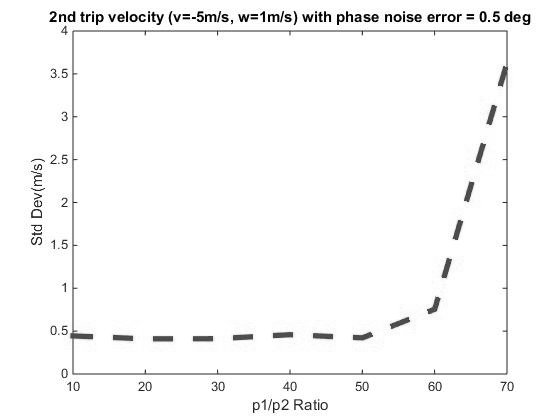}
	\caption{The standard deviation of the second trip velocity retrieval, after frequency planning of $\omega_{1}$ and $\omega_{2}$. }
	\label{fig_IFfig7}
\end{figure}

\subsection{Velocity and Spectral Width Retrieval:}
This scheme gives immense benefit in terms of suppression of the echoes of undesired trip, but it requires spectral processing to retrieve the velocity and spectral width information. This is because, the two frequencies used, make the data samples in adjacent pulses, uncorrelated. If we retrieve the velocity and spectral width, using alternate samples, then the unambiguous velocity range would becomes half. In this section, we highlight a new method, using which we can still recover the original range of velocity, with some constraints.\par
\begin{figure}[!t]
	\centering
	\includegraphics[width=3.8in]{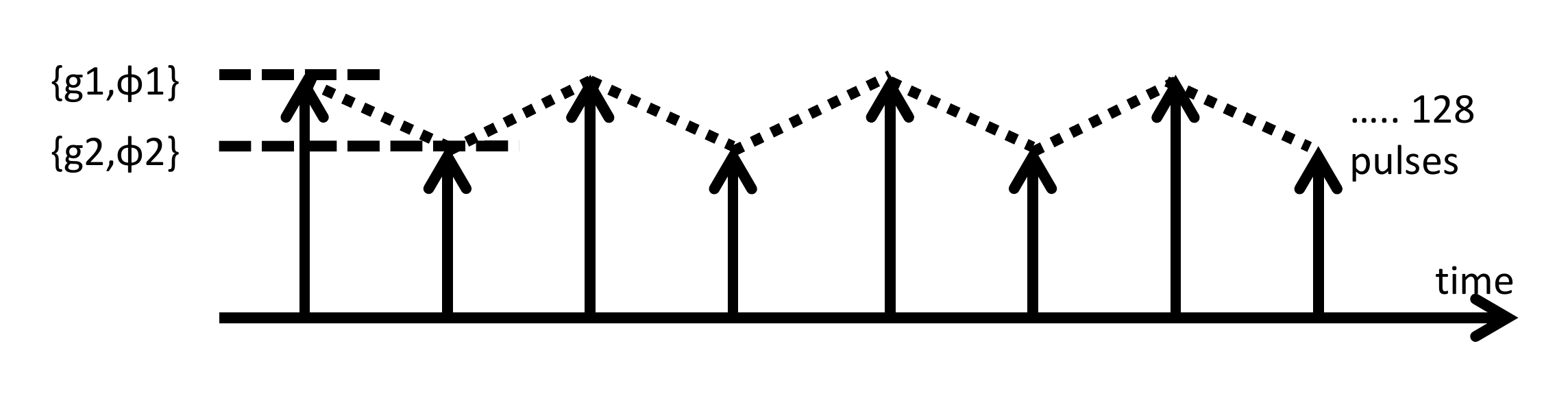}
	\caption{Fixed gain and phase modulation due to uncorrelated frequencies in alternate pulses. }
	\label{fig_GainPaseImb}
\end{figure}

The uncorrelated data from two frequencies, manifest itself with a different gain and phase term, in adjacent pulse. This phase term is in addition to the phase modulation term, due to Doppler. Hence, even for a stationary target, the gain and phase, would go through two states, over coherent integration time. There would be a fixed amplitude and phase modulation with a cycle rate of $F_{PRF}$, and spectrum would have another sideband at $V_{1 or 2} - \pi$. This is illustrated in Fig. \ref{fig_GainPaseImb}. Moreover, both the original and sideband spectrum would look identical and there arises a necessity for another mechanism, to figure out the original velocity. For correct spectral width retrieval, the sideband would need to be filtered out, otherwise there would be over-estimation. We propose a method here to correctly estimate velocity and spectral width, with this type of fixed phase modulation. This can be used under narrow spectral width assumption and we would define a narrow spectral width signal to be one-tenth of the unambiguous velocity range. For S-band, it would be 5 m/s and for Ku band, it would be close to 2 m/s. \par

The proposed method rely on spectral processing to retrieve velocity and spectral width in combination with pulse pair after sideband removal. Another explanation of generation of sideband under fixed amplitude and phase imbalance over the integration pulses is explained next. Under these circumstances, the lag-1 auto-correlation function will be zero and at lag-2 will be one. Hence the auto-correlation function at various lags can be written as:
\begin{equation}
R_{n}^{com} = R_{n}[1 0 1 0 ... 0]
\end{equation} 

where $R_{n}$ is the single lag auto-correlation function. If we take the fourier transform of $R_{n}^{com}$, we get:
\begin{equation}
\begin{aligned}
FT\{R_{n}^{com}\} &= FT\{R_{n}[1 0 1 0 ... 0]\} \\
				  &= FT\{R_{n}\} \ast FT\{[1 0 1 0 ... 0]\} 	
\end{aligned}
\end{equation} 
where $\ast$ is the convolution operator. The term to the right of the convolution operator is the fourier transform of a periodic pulse train. For this, the fourier transform is also periodic and the impulses are spaced by $2pi/N$ (\cite{Oppenheim:2009:DSP:1795494}) with periodicity $N = 2$. Hence the fourier transform of the overall auto-correlation function is the power spectral density of the weather echoes convolved with an impulse train spaced apart by $pi$ radians. Thus now it is easy to understand that the spectrum of weather echoes with odd and even pulses modulated at different frequency will have a sideband at $V_{1/2} - pi$ within the nyquist interval. Now we explain next how to get rid of this sideband.

\subsubsection{Method} \label{VelRet}
\begin{figure}[!t]
	\centering
	\includegraphics[width=3.6in]{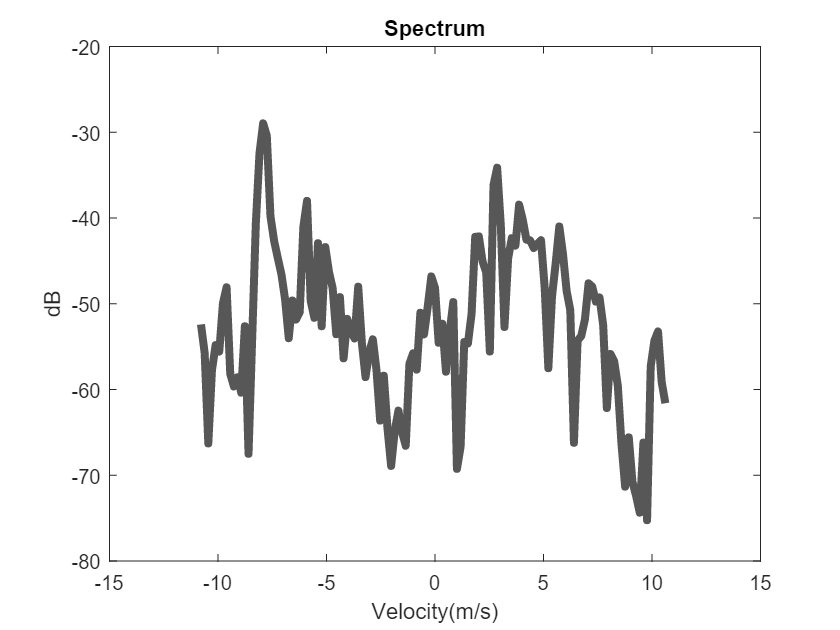}
	\caption{The Velocity Spectrum of one range cell at a certain radial, from D3R weather radar, after frequency planning of $\omega_{1}$ and $\omega_{2}$ in adjacent pulses. The number of pulses considered are 128.}
	\label{fig_d3rSpec}
\end{figure}

We would start with upper-half of the spectrum, of a range gate with $SNR > 10dB$, for a weather echo, within a ray. Assumption is that, either the original or side-band would fall in this region of the spectrum. This is a fair assumption, because the original and sideband spectrum are separated by $\pi$ radian. We run pulse-pair auto-correlation algorithm, on this half of the spectrum (making the other half zero), to get the initial crude estimate of velocity. As a next step, we use a notch filter with normalized notch width equal to 0.5, on the original spectrum, and passband centered around the estimate of velocity, from last step. This would notch out the other component. We run pulse-pair auto-correlation algorithm, once again, to get an accurate estimate of velocity and spectral width. An example spectra from a D3R ray is shown in figure \ref{fig_d3rSpec}.

\begin{figure}[!t]
	\centering
	\includegraphics[width=2.6in]{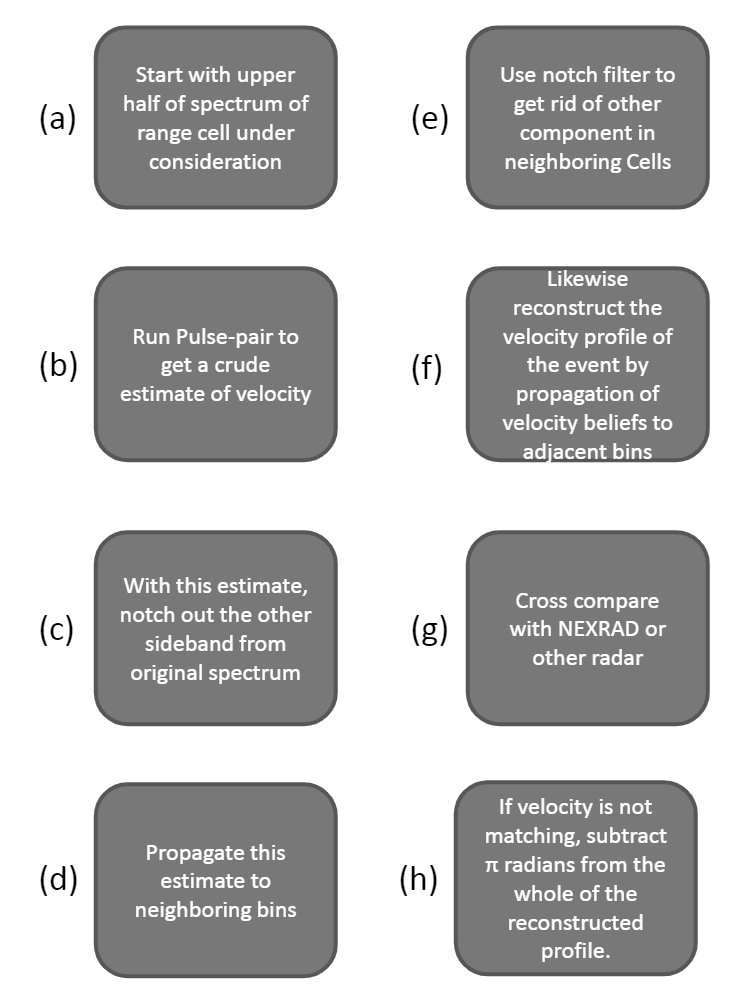}
	\caption{Steps for retrieval of velocity and Spectral width with a frequency diversity scheme.}
	\label{fig_method}
\end{figure}

After the estimate of velocity has been obtained in one range cell, this can be propagated to other neighboring range cells below and above it, and also to the one left and right of it, as an initial crude estimate. Assumption here is that of spatial contiguity of weather signals, which is fair enough for most weather events . With the crude estimate, the notch passband can be centered around the velocity spectrum in the adjoining range bins and the sidebands can be notched out. In turn, again the information of velocity is passed on to the neighboring bins and we continue to get a better estimate of velocity and spectral width, progressively in the same and adjoining rays. However, we need to verify our original assumption of retaining the upper-half of the spectrum, in the very first range cell, that we started from. With this assumption only, the velocity profile was constructed in other range cells. This verification process can be accomplished by comparison with other radars or with different band in the same radar. If we observe that the velocities are not matching, then we need to subtract out $v_{unb}$ from our computed velocities. This would construct a velocity profile, if we had started with the other sideband in the first place. This method is summarized in figure \ref{fig_method}.

\begin{figure*}[!t]
	\centering
	\includegraphics[width=7in]{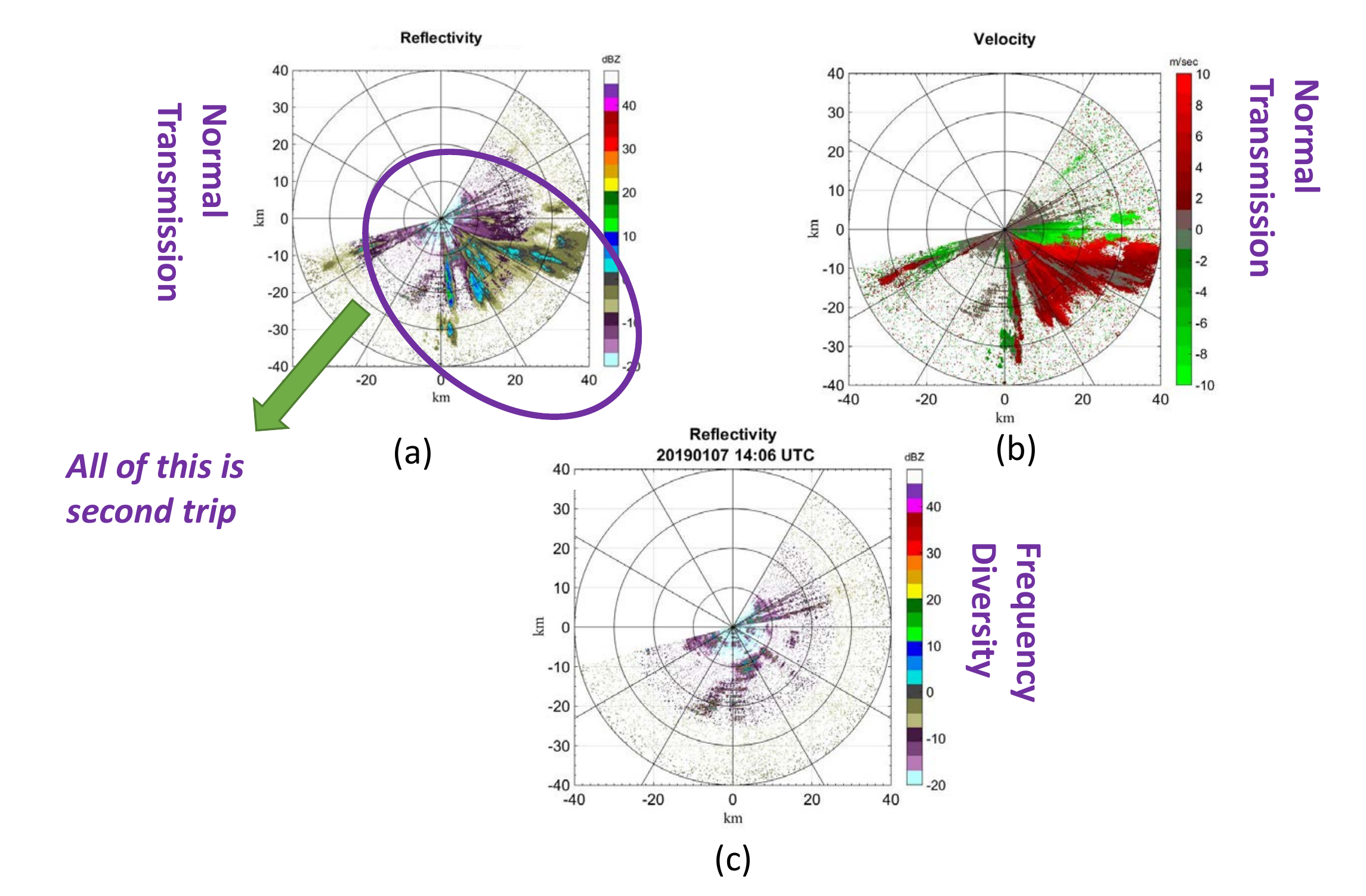}
	\caption{(a) and (b) depict the reflectivity with normal transmission and with frequency change pulse to pulse.  The south-east region has second trips as confirmed with Nexrad radar. There is no first trip in this case. (c) depicts the velocity using normal transmission. The elevation is 2 deg and clearly suppression can be observed for second trip echoes using frequency diversity.}
	\label{fig_case1}
\end{figure*}

\begin{figure*}[!t]
	\centering
	\includegraphics[width=7.5in]{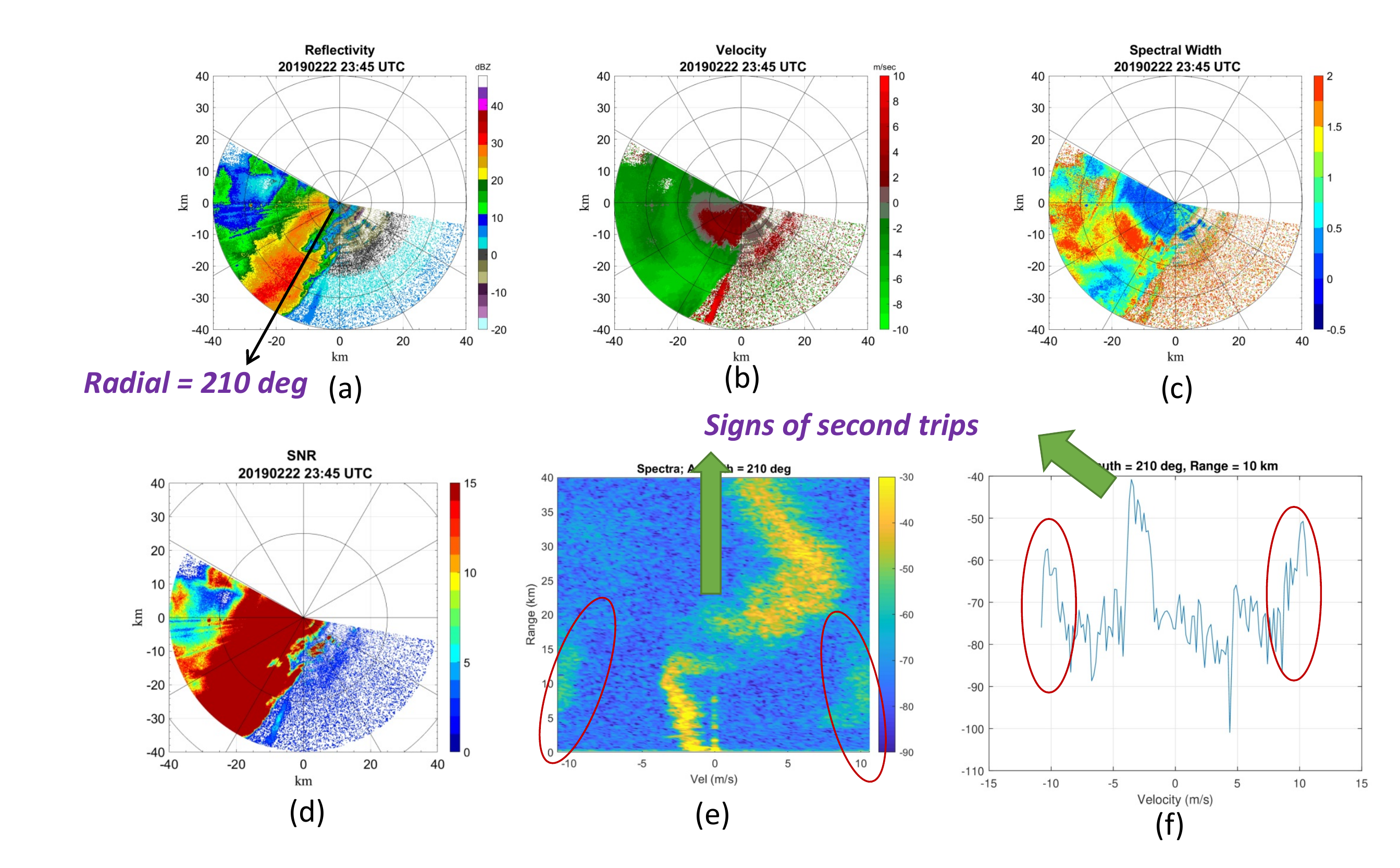}
	
	\caption{(a), (b) and (c) are the reflectivity, velocity and spectral width for an event observed by D3R  with normal transmission scheme. This event had dominant first trip and few traces of second trip in the near range. The velocity profile along a certain ray is shown in (e) with traces of second trip in the 5 to 15 kms of range. Thus, bimodal Gaussian distribution is observed at a range bin at 10kms of range which is plotted in (f).}
	\label{fig_case2}
\end{figure*}

\begin{figure*}[!t]
	\centering
		\includegraphics[width=7.5in]{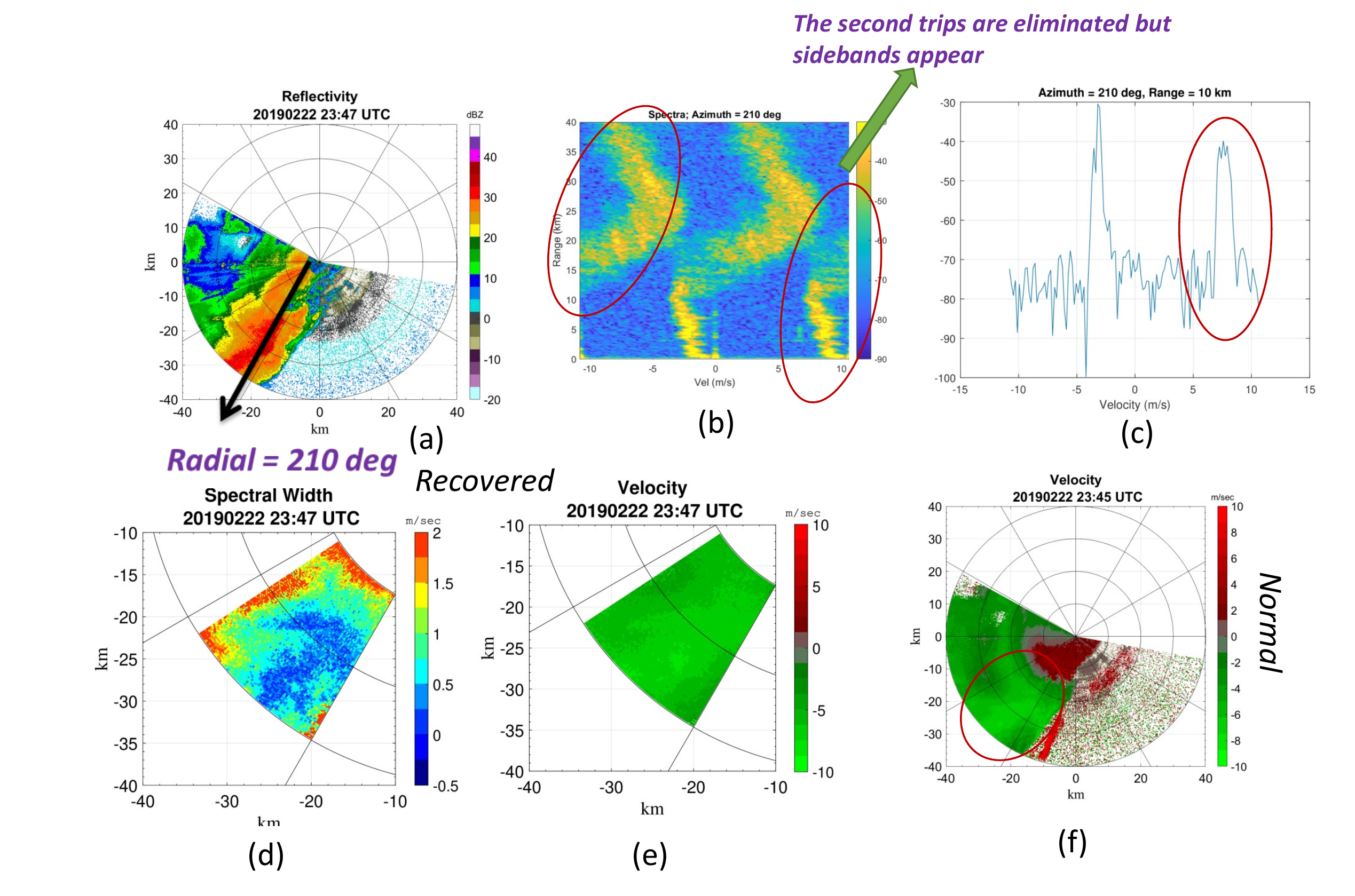}
	
	\caption{(a), (b) and (c) shows reflectivity and velocity profile (for frequency diversity scheme) showing the sideband which occurs due to the gain and phase terms going through two set of states, inducing a modulation over the integration period. (d) and (e) shows recovered velocity and spectral width after removal of the sideband, under narrow spectral width assumption.}
	\label{fig_case3}
\end{figure*}

\begin{figure*}[!t]
	\centering
		\includegraphics[width=5in]{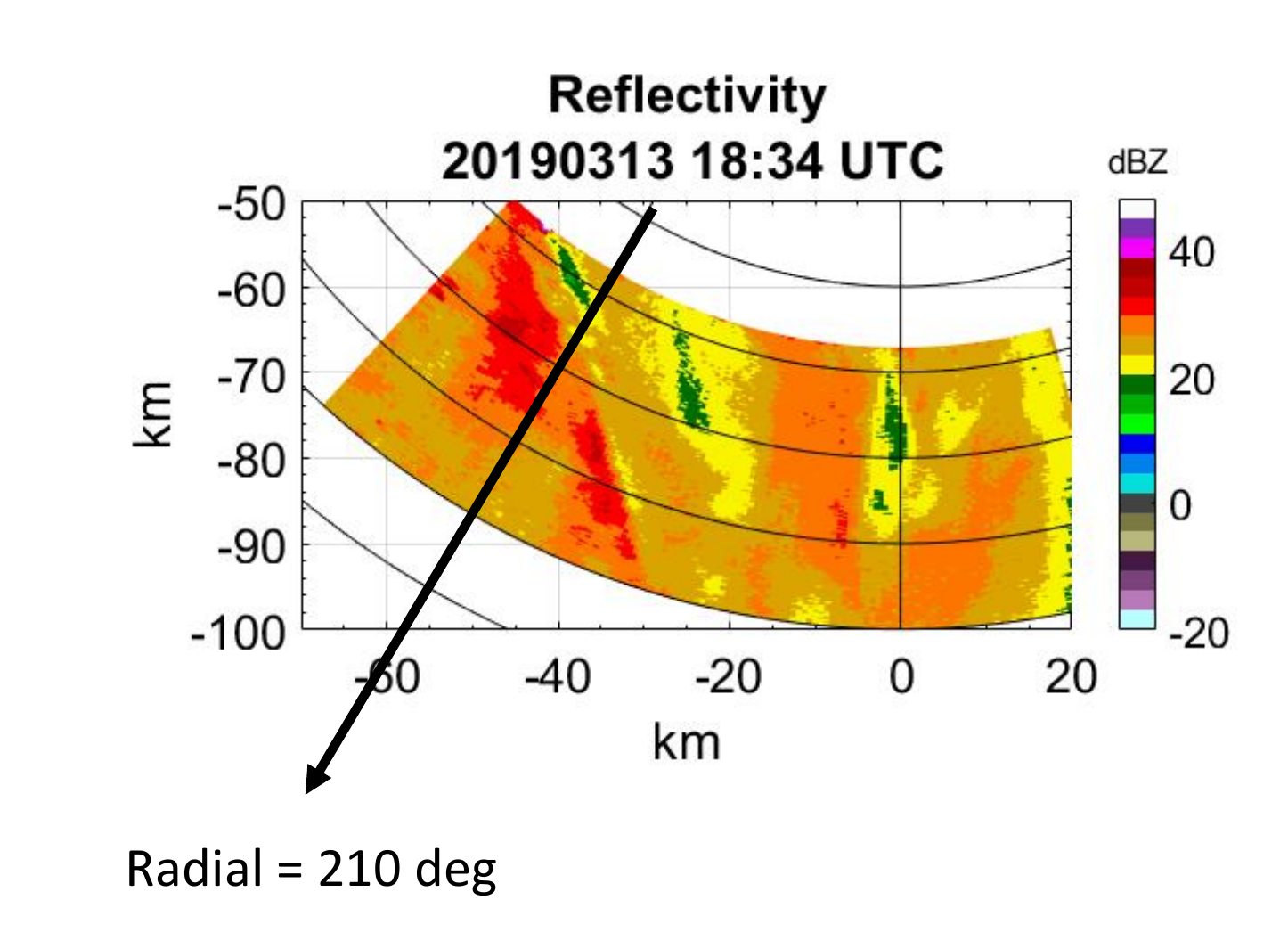}
	\caption{These are the recovered Second trips and plotted with appropriate range. The ppi in (a) has an elevation of 1 deg while that in (b) is 2 degrees.}
	\label{fig_case4}
\end{figure*}

\section{Effect of Frequency Diversity scheme on other dual-polarization moments}\label{section2}
It can be easily observed that the frequency diversity scheme would improve the bias induced by unwanted trip, by its suppression to a level, below noise. Practically it is seen that, the accuracy of dual-polarization moments depend upon the co-polar correlation between the horizontal and vertical polarization signals. \cite{Bringi2001} describe in detail, the effect of co-polar correlation on these moments under alternate and hybrid modes of operation. In this section, we try to analyze the effect of frequency diverse scheme on the estimation of co-polar correlation coefficient. During this, we would also try to see the effect of non-ideal conditions and mis-matched channels. Assume the first trip, for all pulses, be denoted by $H_{1}$ and the second trip by $H_{2}$ and the baseband filter matrix by $\textbf{F}_{bb}$, then the equivalent signal model for H-pol and V-pol, for the first trip retrieval, can be written as:

\begin{equation}
\begin{aligned}
&\textbf{H}^{1} = \textbf{H}_{1} + \textbf{F}_{bbh}.\textbf{H}_{2} \\
&\textbf{V}^{1} = \textbf{V}_{1} + \textbf{F}_{bbv}.\textbf{V}_{2}
\end{aligned}
\end{equation} 
The auto-correlation function for the first trip, for the hybrid mode of operation, can be written as:
\begin{equation}
\begin{aligned}
R^{1}_{vh}(0) &= \frac{1}{N}Tr\{\textbf{V}^{1}\textbf{H}^{1H}\} \\
			  &= \frac{1}{N}Tr\{(\textbf{V}_{1} + \textbf{F}_{bbv}.\textbf{V}_{2})(\textbf{H}_{1} + \textbf{F}_{bbh}.\textbf{H}_{2})^{H}\} \\
			  &= \frac{1}{N}Tr\{\textbf{V}_{1}\textbf{H}_{1}^{H} + \textbf{V}_{2}\textbf{H}_{2}^{H}\textbf{F}_{bbh}^{H}\textbf{F}_{bbv}\}
\end{aligned}
\end{equation} 
If the characteristics of both H and V pol filters is the same, then $\textbf{F}_{bbh} = \textbf{F}_{bbv} = \textbf{F}$ and the above equation could be simplified to:

\begin{equation}
R^{1}_{vh}(0) = \frac{1}{N}Tr\{\textbf{V}_{1}\textbf{H}_{1}^{H}\} + Tr\{\textbf{V}_{2}\textbf{H}_{2}^{H}\textbf{F}^{H}\textbf{F}\}
\end{equation} 
Similarly, for the second trip, we can model it as:
\begin{equation}
\begin{aligned}
&\textbf{H}^{2} = \textbf{F}_{bbh}.\textbf{H}_{1} + \textbf{H}_{2} \\
&\textbf{V}^{2} = \textbf{F}_{bbv}.\textbf{V}_{1} + \textbf{V}_{2}
\end{aligned}
\end{equation} 
and the autocorrelation function:
\begin{equation}
R^{2}_{vh}(0) = \frac{1}{N}Tr\{\textbf{V}_{2}\textbf{H}_{2}^{H}\} + Tr\{\textbf{V}_{1}\textbf{H}_{1}^{H}\textbf{F}^{H}\textbf{F}\}
\end{equation} 
The corresponding Correlation coefficients could be written as \cite{Bringi2001}:
\begin{equation}
\begin{aligned}
&\rho^{1}_{vh}(0) = \dfrac{R^{1}_{vh}(0)}{\sqrt{P^{1h}_{co}P^{1v}_{co}}} \\
&\rho^{2}_{vh}(0) = \dfrac{R^{2}_{vh}(0)}{\sqrt{P^{2h}_{co}P^{2v}_{co}}}
\end{aligned}
\end{equation} 
where $P^{1,2,h,v}_{co}$ is the co-polar power for first or second trip echoes, and for H or V pol channels.
The degree of dissimilarity between the autocorrelations of H and V Pol, will be a factor which would impact the $\rho_{vh}(0)$ for the first and second trip echoes. This dissimilarity could arise due to slight difference in filter characteristics, on the receive (cumulative effects of anti-aliasing or baseband CIC/FIR filter). Now, we would analyze the effect of filter, on differential refkectivity ($Z_{dr}$), when first trip is being retrieved:

\begin{equation}
\begin{aligned}
Z_{dr}^{1} &= \frac{P^{1h}_{co}}{P^{1v}_{co}} = \frac{R^{1}_{vv}(0)}{R^{1}_{hh}(0)} \\
&= \dfrac{Tr\{(\textbf{V}_{1} + \textbf{F}_{bbv}\textbf{V}_{2})(\textbf{V}_{1} + \textbf{F}_{bbv}\textbf{V}_{2})^{H}\} }{Tr\{(\textbf{H}_{1} + \textbf{F}_{bbh}\textbf{H}_{2})(\textbf{H}_{1} + \textbf{F}_{bbh}\textbf{H}_{2})^{H}\} } \\
&= \dfrac{Tr\{\textbf{V}_{1}\textbf{V}_{1}^{H} + \textbf{V}_{1}(\textbf{F}\textbf{V}_{2})^{H} + (\textbf{F}\textbf{V}_{2})\textbf{V}_{1}^{H} + \textbf{F}\textbf{V}_{2}(\textbf{F}\textbf{V}_{2})^{H}\}}{Tr\{\textbf{H}_{1}\textbf{H}_{1}^{H} + \textbf{H}_{1}(\textbf{F}\textbf{H}_{2})^{H} + (\textbf{F}\textbf{H}_{2})\textbf{H}_{1}^{H} + \textbf{F}\textbf{H}_{2}(\textbf{F}\textbf{H}_{2})^{H}\}}
\end{aligned}
\end{equation}

assuming $\textbf{F}_{bbh} = \textbf{F}_{bbv} = \textbf{F}$. It can be easily observed from the equation above, that major contribution towards bias of $Z_{dr}$, is through the middle two terms in numerator and denominator (getting multiplied by the first trip voltage). The second trip voltage, however, is always preceded by filter and is definitely going to be low. Additionally, the degree of dissimilarity between the filter response on the H and V Pol channels, is also going to contribute towards bias in $Z_{dr}$.

\section{Performance test and validation on D3R}\label{section3}
With D3R weather radar, we are able to make co-aligned Ku and Ka band observations for a precipitation event. It is a very useful ground validation tool for Global Precipitation Measurement mission (GPM) satellite with dual-frequency radar. D3R uses a combination of short and medium pulses with pulse duration of $1 \mu s$ and $20 \mu s$ respectively. The short pulse is used to provide adequate sensitivity for the duration of medium pulse and mitigate blind range of the medium pulse. The radar has been in numerous field campaigns (see \cite{First5yrs}, \cite{FiveYrsOp}, \cite{olempex1}) Recently, the D3R radar was upgraded with a new version
of digital receiver hardware and firmware which supports
larger filter length and multiple phase coded waveforms, change of frequencies pulse to pulse and also newer IF sub-systems (\cite{Kumar8517944}, \cite{8128188} and \cite{AdapFilt}, \cite{kumar2019receive}). With these newer sub-systems, D3R was deployed for observing snow at the winter Olympics in pheongchang region of South Korea, 2018 (\cite{8899120}, \cite{Icepop1}). With a $500 \mu s$ pri, D3R's unambiguous range is 60 kms, but the first $130 \mu s$ is used to inject noise, for receiver calibration. Hence, D3R first trip is upto 40 kms with a dead range of 20 kms. Beyond 60 kms, it is the second trip range. In this section, we would demonstrate the frequency diversity scheme, with a pulse by pulse change of frequency, on D3R Ku band, and also the retrieval of second trip after 60 kms of range. The first case that is demonstrated in figure \ref{fig_case1}, has all of second trip echoes and no weather echoes in the first trip range. The normal transmission is using a chirp waveform for medium pulse centered at 65 MHz and a short pulse at 55 MHz. Whereas, for frequency diversity case, the frequencies used are 55 and 65 MHz for short and medium (odd numbered pulses). For the even numbered pulses, the frequencies are 60 and 70 MHz for short and medium pulse respectively. The suppression of second trip can be observed in the south-east sector. The remaining echoes are the clutter points in near range. Another case is depicted in figure \ref{fig_case2}. Initially, we show normal chirp transmission, as reference, with information about velocity and spectral width. This case has first trip in the south-west sector, with second trip power overlaid. Figure \ref{fig_fifth_case1} shows the velocity profile along 210 radial and clearly the second trip contamination can be observed in 5 to 15 kms of range. Also, figure \ref{fig_sixth_case1} plots the spectrum at range 10 km and at radial $210 \degree$ azimuth, showing the second trip velocity. The frequency diversity case is shown in figure \ref{fig_case3}, which is taken a couple of minutes later, than the normal transmission. There were no second trip signatures from 5 to 15 kms of range at the same radial but instead there is a replica of the original velocity spectrum as sideband, appearing at a distance of $\pi$ radians from the original. Processing to remove this sideband, has been explained in section \ref{VelRet}. After we go through the steps listed there, we can reconstruct velocity and spectral width through filtering out sideband power. The velocity and spectral width recovered after this process is shown in figure \ref{fig_fourth_case2} and \ref{fig_fifth_case2} respectively.\par
Also, for this case, we have recovered second trip, which is shown after 60 km range in figure \ref{fig_case4}. For doing first trip retrieval, we programmed sequence, $\omega_{1}$, $\omega_{2}$, in a frame, while for second trip recovery, the sequence of  $\omega_{2}$, $\omega_{1}$, was used by the NCO. Short pulse sub-channel was configured to recover second trip and the medium pulse sub-channel for first trip. Hence, both trips were being retrieved simultaneously. However, we would no longer have the short pulse echoes for the first trip, which mitigates the blind range of the medium pulse. Due to this, there is a gap in the beginning (first $\sim 4$ km in ppi) and a bigger gap can be observed in between first and second trip echoes (in addition to the dead range).

\section{Conclusion} \label{section5}
We developed the scheme of Inter pulse frequency diversity techniques for weather radar systems and utilized the orthogonality between two frequencies, in IF domain, to reject out the undesired trip echoes. This technique shows improvement in performance of second trip suppression and retrieval under phase noise condition, compared with Chu phase codes (SZ Codes). Extensive time-series simulations were carried out to ascertain the performance of this technique. Comparison with Chu phase code based inter-pulse system was presented and this shows promising results with recovery of the weaker trip echoes under wider dynamic range of overlaid power contamination.\par
However, it should be emphasized that due to the un-correlated data samples in adjacent pulses, velocity and spectral width needs to be reconstructed from coherent processing interval, with a new method, that is described. But it would work under assumption of narrow spectral width. Such assumptions also holds for SZ code based retrievals, which under wider spectral width tend to behave more like random phase codes.

\ifCLASSOPTIONcaptionsoff
  \newpage
\fi

\bibliographystyle{IEEEtran}
\bibliography{references_arxiv1}

\end{document}